# Measurement of the Phase Diagram of DNA Unzipping in the Temperature- Force Plane.

C.Danilowicz, Y. Kafri, R.S. Conroy, V. W. Coljee, J. Weeks, M. Prentiss

**Physics Department, Harvard University, Cambridge, MA 02138, USA.**

Contact Details:

Tel. +1-617-495-4483   Fax. +1-617-495-0416   e-mail: prentiss@fas.harvard.edu

## Abstract

We separate double stranded lambda phage DNA by applying a fixed force at a constant temperature ranging from 15°C to 50°C, and measure the minimum force required to separate the two strands, providing the first experimental determination of the phase boundary between single stranded DNA and double stranded DNA in the temperature-force plane. The measurements also offer information on the free energy of dsDNA at temperatures where dsDNA does not thermally denature in the absence of force. While parts of the phase diagram can be explained using existing models and free energy parameters, others deviate significantly. Possible reasons for the deviations between theory and experiment are considered.

Studies of the mechanical separation of double stranded DNA (dsDNA) enhance the understanding of DNA replication in vivo. Although it is possible to separate dsDNA by heating well above body temperature in procedures such as the polymerase chain reaction (PCR), in living organisms DNA replication is a very complex process assisted by a variety of specialized proteins. DNA unzipping, the separation of dsDNA by applying a force that pulls two strands of DNA apart at one end of the molecule, is similar to some of the steps occurring *in vivo*. Several techniques such as optical and magnetic tweezers, and atomic force microscopy have been used in single molecule studies of the unzipping of DNA and the unfolding of RNA [1,2,3,4,5,6,7,8,9].

Recent theoretical work focused on the unzipping of DNA under a constant applied force, where both the unzipping of homopolymers and heteropolymers has been considered [10,11,12,13,14,15,16]. These models assume thermal equilibrium and predict that unzipping is a first order phase transition with a minimum required force that decreases with increasing temperature, ending at the thermal denaturation point where no force is required to separate the two strands [17]. Theoretical frameworks that successfully encompass both the traditional thermal separation and the recently investigated force-induced separation should enhance the predictive power of models of DNA binding potentials that may improve the design of PCR primers and DNA chips.

In such models, the phase diagram in the temperature-force plane is obtained by comparing the free energy of the unzipped molecule under an applied constant force to that of force-free dsDNA. Provided that a good estimate for the free energy of single stranded DNA (ssDNA) under an external force is possible, one can then use the critical force to estimate the free energy of dsDNA.

Previous measurements of the unzipping of DNA have been made at room temperature [7,8]. In this paper we evaluate the unzipping of the first 1500 of the 48502 base pairs of lambda phage DNA by applying a fixed force at a constant temperature, at temperatures varying between 15°C and 50°C. We show that the force required to unzip the heterogeneous DNA decreases with increasing temperature, in accord with previous predictions [10,11,12,13]. We have also observed the rezipping of dsDNA when the applied force is below 2 pN for an hour or longer. Our results show the first experimental phase diagram for dsDNA separation at constant applied force and suggest that a) there is reasonable agreement with projections from bulk thermodynamics between 25°C and 35°C; b) the occurrence of hairpins in the ssDNA neglected in theoretical models, may play an important role at high temperature; c) the temperature dependence of the free energy of the dsDNA changes in a non trivial manner at low temperatures.

The critical force required to unzip dsDNA can be calculated using the expression for the free energy difference per base pair, $\Delta G$, between a base pair in dsDNA and the same base pair in a ssDNA stretched at the ends by an external force, F [11,12]. This free energy difference is given by

$$\Delta G = g_b - 2\, g_u\, (F) \qquad (1)$$

where $g_b$ is the average free energy per base pair of double stranded DNA (at zero force) and $g_u$ (F) is the free energy per base pair for each of the stretched single strands. The value of $g_b$ depends on the specific sequence of the dsDNA and is usually estimated using a nearest neighbor approximation [18]. The dsDNA will separate into ssDNA when the applied force exceeds the critical force $F_c$ such that

$$g_b = 2\ g_u\ (F_c) \qquad (2)$$

Though much useful information can be gained from models that treat dsDNA as a homopolymer [10,11,12,13,14], DNA contains a well-defined sequence of base pairs making it a heteropolymer rather than a homopolymer. For homopolymers, the free energy difference between the stretched ssDNA and the dsDNA is the same for all of the base pairs in the sequence, and the free energy difference is a linear function of the number of unzipped base pairs [19]; therefore, a constant applied force, larger than the critical force, $F_c$ (T), will unzip homopolymeric dsDNA at a constant rate. In contrast, in heteropolymeric dsDNA, the free energy difference between the stretched ssDNA and the dsDNA does not have the same value for all base pairs. As a result, the energy landscape for this heteropolymer displays a significant sequence dependent variation as a function of the number of unzipped base pairs [11,12,20]. As discussed in Refs. [11,12,20], for a wide range of forces near the critical force the dsDNA will unzip in a series of different, discrete, sequence dependent steps. The locations in the sequence where the unzipping pauses, have been successfully predicted by Monte-Carlo simulations [21]. The critical force $F_c$ for a given sequence of dsDNA is the minimum force required to unzip the entire sequence. This unzipping in discrete jumps near $F_c$ makes the determination of the critical force difficult, and forces us to define some procedure for locating it (see below).

We measure the unzipping of lambda phage DNA using a massively parallel system [20,22] and DNA samples prepared using the techniques described in detail in Ref. [20]. A schematic representation of the apparatus is shown in Fig. 1. The molecular construction consists of two covalently linked lambda phage dsDNA molecules. The dsDNA attached to the surface is a spacer that separates the dsDNA to be unzipped from the surface to avoid surface interactions. The spacer is labeled with digoxigenin and therefore can be specifically attached to a glass surface through an antigen-antibody interaction. The second DNA molecule is closed at one end with a hairpin, and the other end has a biotin label that can be bound to a streptavidin coated superparamagnetic bead. Finally, the spacer attached to the glass surface can be stretched to its contour length of 16.5 μm by applying a small force (2 pN) before the unzipping experiment begins at higher force. Before starting the experiment, the sample containing the DNA and the beads is placed in a 0.8 mm clean square capillary with a 0.55 mm round inner capillary and incubated. The temperature of the square capillary can be varied from 10 to 80°C using a thermoelectric cooler placed on top of the aluminum mount holding the square capillary. A temperature sensor close to the capillary channel provides feedback for the stabilization loop controlling the thermoelectric cooler.

The inner round capillary is modified with anti-digoxigenin antibody by an overnight adsorption at room temperature and can therefore specifically interact with the DNA construction. The optical apparatus, which consists of an inverted microscope with a computer-controlled frame grabber, tracks the position of the superparamagnetic beads as a function of time for each different force and temperature condition. The distance $D$ between the bead and the inner glass capillary depends both on the number of base pairs that are unzipped and the extent to which the unzipped ssDNA and the dsDNA linker are stretched. The stretching of both dsDNA and ssDNA depends on the applied force [23,24,25,26], so a distance $D$ will correspond to different numbers of unzipped base pairs when different forces are applied to the magnetic bead [20]. The stretching behavior of dsDNA shows that it is relatively stiff at forces above 10 pN while ssDNA exhibits a higher flexibility. The average length of a dsDNA spacer under 10 pN of force is 16.2 μm, and the average length under 15 pN of force is 16.3 μm. Similarly, at an applied force of 10 pN the average total length of the dsDNA space plus a completely unzipped lambda phage ssDNA is 65 μm whereas at an applied force of 15 pN the total average length is 77 μm [27]. Thus, we can calculate from the stretched length of the ssDNA at 10 pN, an extension per base pair of 0.5 nm, and 0.63 nm at 15 pN. These values have also been confirmed with a separate measurement of lambda phage ssDNA obtained through melting at 99°C in a thermal cycler and cooling down quickly to 5°C; an average of 5 beads showed an extension of 24.2 μm and 27.8 μm at 10 pN and 15 pN, respectively.

Initially, when the magnet approaches the square channel and a low force is applied (3-5 pN), the beads separate from the surface becoming tethered at a distance of 16- 17 μm. At the desired temperature (between 18°C and 50°C), the measurement starts when the force is increased to a value between 3 to 25 pN whereas the force range for the experiment at 15°C was 18 to 55 pN. The displacement of the beads is followed in time and the distance to the surface can be measured precisely. Typically, the concentration of DNA is several times smaller than the concentration of the beads so that each bead is connected through one single DNA molecular construction.

Fig. 2 shows the measured phase diagram as a function of force and temperature. The red circles in the figure represent data from samples that were initially incubated at 50°C for 30 minutes and were then equilibrated to the final temperature at which the phase diagram was measured. The sample was allowed to equilibrate at each temperature for 20 minutes before the experiment started. Samples that were initially at 50°C will be referred to as hot samples. The green squares show the force required to unzip the dsDNA as a function of temperature for samples that were left at each unzipping temperature for more than 15 hours before unzipping, so they should be approaching the equilibrium value for the critical force. In order to find the force at which the phase transition occurs for a given temperature, we conducted a series of experiments on the same group of single molecules where we applied an initial force $F_o$ to the sample for 15 minutes, and then measured the number of unzipped molecules by the end of that time. A molecule was considered to have begun to unzip if the distance between the bead and the surface increased suddenly by more than 1.5 µm (1500 base pairs) during the time interval, and the bead remained tethered for more than 1 minute after the sudden increase in the distance from the surface. The force was then increased to $F_1$, and the molecules were allowed to unzip for another 15 minutes. We counted the number of molecules that had remained zipped when the applied force was $F_o$, but began to unzip when the force was $F_1$. We then increased the force to $F_2$, and repeated the process. As a final step we increased the force to 25 pN for 30 minutes, and measured the total number of molecules that began to unzip. We assumed that at such a large applied force all molecules that were correctly tethered to the surface would begin to unzip, and that any beads that remained tethered must have been bound to the surface by an incorrect construction. We then calculated the fraction of the correctly bound molecules that unzipped at or below a given average applied force. There is a variation in the magnetization of the beads, so if the average force on the beads is the critical force, half of the molecules should unzip and half should remain zipped; therefore we defined the measured critical force as the value of the average force at which 50% of the correctly bound molecules begin to unzip at a given temperature. The critical forces measured using this method, were quite reproducible from sample to sample; each experiment was done at least twice, and some experiments were done several times with samples prepared separately. The numbers of correctly tethered dsDNA molecules in a given experiment ranged from 20 to 50, with an average value of approximately 30, so many single molecule measurements were included in each experiment. The variation in average force between different samples under identical conditions was less than 1 pN. The temperature inside the square channel was measured at the end of each experiment. We also conducted a few experiments where we applied a single constant force for an hour, and checked that 50% of the correctly bound molecules did unzip at the critical force determined using the method described above allowing us to verify that the 15 minute -interval was adequate to sample each applied force.

The purple diamond in Fig.2 shows the measured melting temperature, $T_m$, for lambda phage dsDNA in the same phosphate saline buffer used in the unzipping experiments, where the melting temperature was determined by a bulk measurement of circular dichroism spectrum as a function of temperature. This denaturation temperature required to unzip the DNA at zero force is in good agreement with theoretical calculations reported for similar ionic conditions [28].

It is interesting to compare the experimental results to a simple theory that utilizes free energy estimates for both the unzipped ssDNA, $g_u$, and the bound DNA, $g_b$. At forces above 10 pN, the measured force versus extension curves for ssDNA have been well described by the mFJC model [29,30]. Below these forces hairpins begin to play an important role and the model is no longer valid. If one measures the free energy of the stretched ssDNA with respect to the unstretched ssDNA (so that $g_F(F=0)=0$), then within the mFJC model the free-energy per base pair of the unzipped DNA is:

$$g_u = -\frac{l}{b} k_B T \ln\left(\frac{k_B T}{Fb} \sinh\left(\frac{Fb}{k_B T}\right)\right) + \frac{l}{2b} \frac{(F\ell)^2}{k_B T} \qquad (3)$$

where $l$ is the distance between base pairs (we use a base pair separation, $l$ = 0.7 nm [1] which predicts an extension per base pair of 0.54 nm at 10 pN and 0.58 nm at 15 pN, (consistent with the length per base pair vs force that we measured for the ssDNA stretched by forces between 10 and 15 pN and with the spacing between phosphates for C2'-*endo* pucker in ssDNA), b is the Kuhn length of ssDNA (current estimate b = 1.9 nm) and $\ell$ = 0.1 nm is a length which characterizes the elasticity of a bond along the ssDNA between bases. The mFJC does not include the possibility of hairpin formation and is not expected to correctly describe the free energy of the ssDNA at forces below 10 pN where hairpins can occur in ssDNA [29].

Given that the free energy reference point was chosen so that $g_u(F=0)$=0, the free energy of the bound dsDNA is just the free energy difference between dsDNA and ssDNA and can be expressed in terms of ΔH and ΔS, the difference in enthalpy and entropy between dsDNA and ssDNA. Then:

$$g_b = \Delta H - T\Delta S \qquad\qquad (4)$$

where T is the temperature in degrees Kelvin. Here the dsDNA is assumed to be completely bound with no denatured loops inside. This assumption is justified to a very good approximation up to the thermal denaturation temperature due to the extremely small Boltzmann weight (or cooperatively parameter) $\sim 10^{-4}$- $10^{-5}$ [31], associated with initiating such a loop. Previously it has been assumed that *ΔH* and *ΔS* are independent of temperature [18]. Given this assumption, at the melting temperature $T_m$, $F_c$=0 and $T_m = \Delta H/\Delta S$ [11,12,18]. The value of ΔS= -20.6 cal/ °K.mol (-1.43 x $10^{-22}$ J/ °K. molecule), is obtained by averaging over the first 1500 base pairs (corresponding to the experimentally measured critical force) using the values of $\Delta$S found in Ref. [18]. The value of $\Delta H = T_m \Delta S$ is then calculated using the experimentally measured thermal denaturation temperature. The value $\Delta H = -7.5$ kcal/ mol (-5.22 x $10^{-20}$ J/ molecule) agrees with modifications of the values of Ref. [18] due to the ionic concentration as described in Ref. [28]. It was assumed that $\Delta$S was not significantly affected by the change in ionic concentrations.

Using the free energy estimates along with Eq.(2), it is then straightforward to calculate numerically the critical force as a function of temperature. The resulting phase diagram is shown as the blue line in Fig. 2. The line is in good agreement with the data in the temperature range from 24°C and 35°C, though there are significant deviations outside this temperature range. The theory dotted does not fit the experimental values at temperatures above 35°C where the critical force is predicted to be below 10 pN. The line is dotted because the mFJC does not accurately predict measured force vs extension curves for ssDNA in this force range, possibly because hairpin formation is neglected [29]. Inclusion of hairpins in the theory would reduce the predicted difference in binding energy between ssDNA and dsDNA, and might bring the theoretical values closer to those observed in the experiments. Finally, recent experiments suggest that bubbles of 2 to 10 bases pairs can form spontaneously in dsDNA at 37°C [32]. Such bubbles would decrease the binding energy of the dsDNA and increase its entropy. Inclusion of these effects would further reduce the predicted critical force making it more consistent with the measured values in the temperature range between 37°C and 50°C.

At temperatures from 15°C to 24°C, the predicted and measured values again diverge. Though the theoretical curve has an approximately constant slope in the temperature range from 15 to 40°C the measured critical force does not. At temperatures below 22°C the measured critical force increases much more steeply with decreasing temperature than it does in the temperature range from 24°C to 35°C, where the theory and experiment are in reasonably good agreement. We attribute the observed decrease in slope at low temperatures to a thermally induced change in the dsDNA conformation. Other signatures of this conformational change include variations in the circular dichroism spectrum and the persistence length of the dsDNA [33]. This hypothesis is supported by preliminary data suggesting that the stretching curves for ssDNA have little dependence on temperature at forces greater than 10 pN. Therefore, $g_u$ is expected to be weakly influenced by temperature.

We have measured the phase diagram for the unzipping of single molecules of lambda phage dsDNA as a function of force and temperature. In the temperature range 24-35°C, the results agree well with projections from bulk thermodynamic data that assume $\Delta S$ and $\Delta H$ are independent of temperature. Above 35°C and below 24°C, the critical force required to unzip dsDNA at a given temperature deviates significantly from calculations based on simple projections of the bulk thermodynamic measurements. At temperatures above 35°C the measured critical force is much smaller than the predicted critical force. This difference may be partly attributed to the formation of hairpins in the ssDNA when the applied force is less than 10 pN. At temperatures below 24°C, the measured critical force is larger than predicted. Temperature dependent conformational changes in dsDNA may contribute to this discrepancy. Thus, the observed phase diagram for the unzipping of lambda phage dsDNA is much richer than earlier theoretical work had suggested. Adding information on DNA conformational changes and hairpin formation could greatly improve the predictive power of theoretical treatments allowing more accurate biological predictions.

# Acknowledgements

We thank David R. Nelson and Maxim D. Frank-Kamenetskii for valuable conversations. This research was funded by grants: MURI: Dept. of the Navy N00014-01-1-0782; Materials Research Science and Engineering Center (MRSEC): NSF # DMR 0213805 and NSF Award #PHY-9876929; NSF grant DMR-0231631. Y.K. acknowledges support from Fulbright Foundation, Israel.

Figure Captions

Fig.1. Schematic of the DNA binding to the inner glass capillary and the magnetic bead. The magnet exerts an attractive force on the superparamagnetic beads pulling them away from the surface at low force. At forces above the critical force the double stranded DNA shown on the left side of the diagram can be separated into two single DNA strands and complete separation is avoided by including a hairpin at the end of the DNA to be unzipped.

Figure 2. Measured phase diagram in the temperature-force plane. The red circles correspond to samples previously incubated at 50°C (hot samples) and the green triangles correspond to samples that were left at each unzipping temperature for more than 15 hours before unzipping. The blue line is calculated according with theory. The purple symbol corresponds to the experimental value for lambda phage DNA melting using circular dichroism in phosphate saline buffer pH 7.4.

**Figure 1**

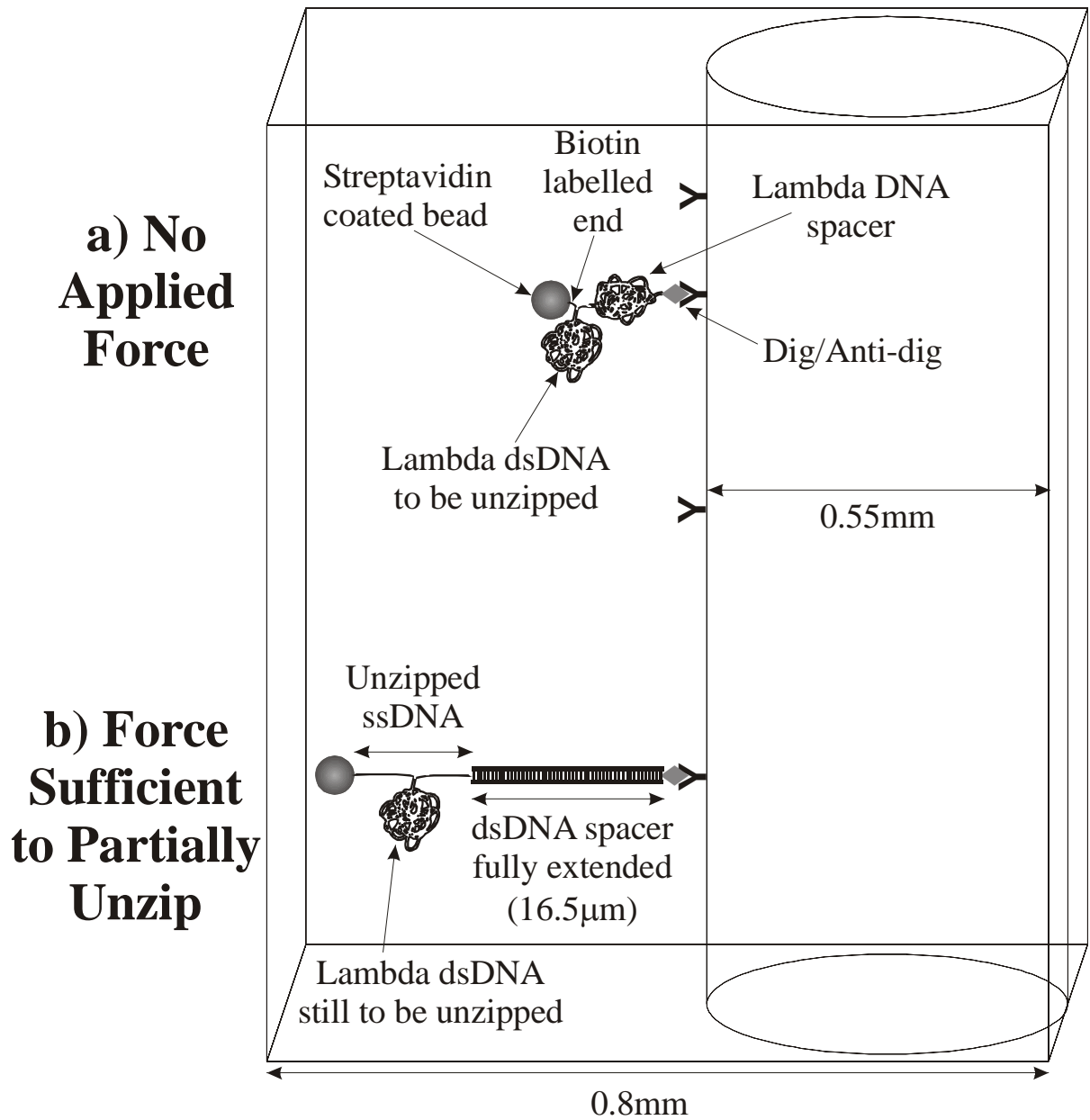

**Figure 2**

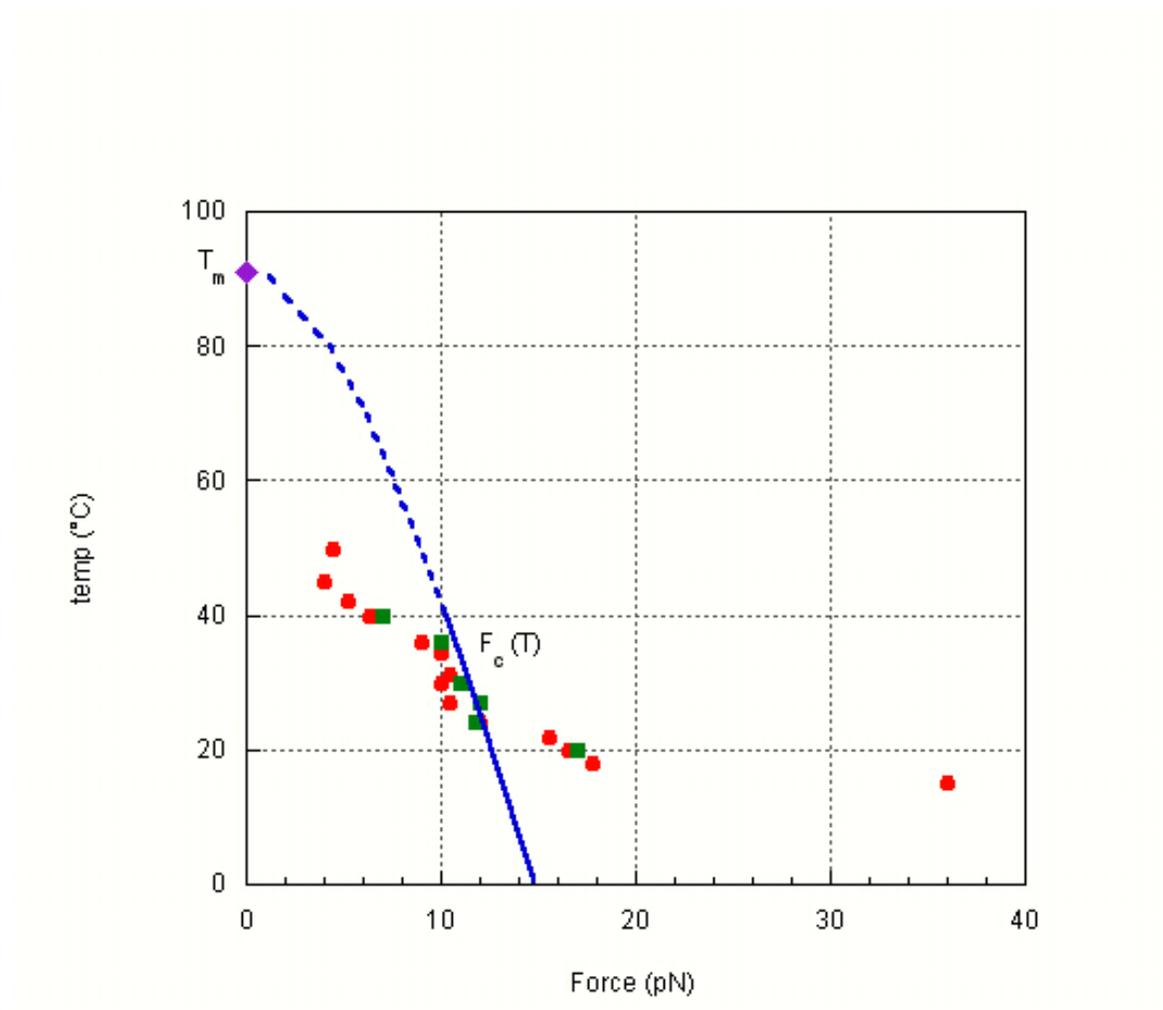